\documentclass[10pt]{iopart}
\usepackage{amssymb,amsmath}
\usepackage[flushleft]{threeparttable}
\usepackage{graphicx,array,multirow}
\usepackage{xspace}%
\usepackage{enumitem}
\usepackage{iopams}  

\begin{document}

\newcommand{\mg}{Mn$_5$Ge$_3$\xspace}
\newcommand{\mgx}{Mn$_5$Ge$_3$C$_x$\xspace} 
\newcommand{\mgcone}{Mn$_5$Ge$_3$C$_{0.1}$\xspace}
\newcommand{\mgctwo}{Mn$_5$Ge$_3$C$_{0.2}$\xspace}

\title{Ferromagnetic resonance and magnetic damping in C-doped \mg}


\author{C-E Dutoit}
\address{Aix-Marseille Universit\'{e}, CNRS, IM2NP UMR7334, F-13397 Marseille
Cedex 20, France}

\author{V.O. Dolocan}
\ead{voicu.dolocan@im2np.fr}
\address{Aix-Marseille Universit\'{e}, CNRS, IM2NP UMR7334, F-13397 Marseille
    Cedex 20, France}

\author{M. D. Kuz'min}
\address{Aix-Marseille Universit\'{e}, CNRS, IM2NP UMR7334, F-13397 Marseille
    Cedex 20, France}

\author{L. Michez}
\address{Aix-Marseille Universit\'{e}, CNRS, CINaM UMR7325, 13288, Marseille, France.}

\author{M. Petit}
\address{Aix-Marseille Universit\'{e}, CNRS, CINaM    UMR7325, 13288, Marseille, France.}

\author{V. Le Thanh}
\address{Aix-Marseille Universit\'{e}, CNRS, CINaM    UMR7325, 13288, Marseille, France.}

\author{B.Pigeau}
\address{Universit\'e Grenoble Alpes, CNRS, Inst. N\'{e}el, F-38042, Grenoble, France}

\author{S.Bertaina}
\address{Aix-Marseille Universit\'{e}, CNRS, IM2NP UMR7334, F-13397 Marseille
    Cedex 20, France}

\begin{abstract}
Ferromagnetic resonance (FMR) was used to investigate the static and dynamic magnetic properties of carbon-doped \mg (C$_{0.1}$ and C$_{0.2}$) thin films grown on Ge(111). The temperature dependence of magnetic anisotropy shows an increased perpendicular magneto-crystalline contribution at 80K with an in-plane easy axis due to the large shape contribution. We find that our samples show a small FMR linewidth (corresponding to an intrinsic magnetic damping parameter $\alpha$=0.005), which is a measure of the spin relaxation and directly related with the magnetic and structural quality of the material. In the perpendicular-to-plane geometry, the FMR linewidth shows a minimum at around 200K for all the samples, which seems to be not correlated to the C-doping. The magnetic relaxation parameters have been determined and indicate the two-magnon scattering as the main extrinsic contribution. We observe a change in the main contribution from scattering centres in \mgctwo at low temperatures, which could be related to the minimum in linewidth.

\end{abstract}

\maketitle
\ioptwocol


\section{INTRODUCTION}

The field of semiconductor spintronics is rapidly developing nowadays. The idea to combine the well established data processing capabilities of semiconductor electronics with ferromagnetism may lead to new functionalities and low power consumption of devices\cite{Zutic,Awschalom}. One of the main obstacle for spin injection into a semiconductor is the conductivity mismatch at the interface of the ferromagnetic metal and the semiconductor\cite{Schmidt}. One way to avoid it is to use a thin insulating layer acting as a tunnel barrier between the two materials. Another approach is to design the spin injecting interface with a similar structure and properties by alloying or doping the semiconductor with a magnetic element.

The intermetallic magnetic \mg could provide the desired solution as it grows directly onto Ge substrate\cite{Zeng2003}, therefore being compatible with existing semiconductor technology. \mg shows ferromagnetism with a Curie temperature (T$_c$) around room temperature\cite{Gajdzik} and an important spin polarization (up to 42\%)\cite{Panguluri,Picozzi}. The \mg hexagonal cell contains 10 Mn atoms which are arranged in two different sublattices (Mn$_I$ and Mn$_{II}$) due to different coordination. Inserting carbon atoms into interstitial voids of Mn$_{II}$ octahedra leads to an increase of T$_c$ up to 450K, supplying a solution for the room temperature spin injection\cite{Surgers}. \textit{Ab-initio} calculations indicate that the structural distortions have a small influence on the increased T$_c$ in \mgx (the lattice is compressed compared to pure \mg), with the enhanced ferromagnetism attributed to a 90$^{\circ}$ ferromagnetic superexchange mediated by carbon\cite{Slipukhina}. 

Several preparation methods were used to grow \mg thin films. The most common growth method is the solid phase epitaxy which consists in the deposition of Mn or Mn and C on a Ge(111) layer followed by an annealing leading to the formation of the \mg or \mgx films. Due to the low Mn solubility in Ge, secondary precipitates or Mn-rich regions/clusters frequently appear inside the \mg films. Mn atoms also diffuse in the underlying Ge(111) substrate which deteriorates the interface quality. In this article, we report on the structural and magnetic properties of thin films C-doped \mg epitaxially grown on Ge(111) by reactive deposition epitaxy (RDE) at room temperature. The low growth temperature reduces segregation and allows the formation of thin films of excellent crystalline quality suitable for the determination of various magnetic parameters by FMR:  magnetic anisotropy, magnetisation and the \textit{g}-factor which were quantitatively determined and theirs dependence on carbon content and temperature was identified. From the study of the FMR linewidth, the magnetic relaxation process is investigated and the magnetic relaxation parameters are found. The main relaxation channels we identify are the intrinsic Gilbert damping and the two-magnon scattering. The intrinsic magnetic damping measured by FMR determines the time scale of the dissipation of magnetic energy into the lattice. We determine that C-doped \mg has a very low damping of spin motions, with an intrinsic Gilbert damping constant $\alpha$ between 0.005 and 0.01, which is one of the lowest known value for ferromagnetic Mn-based thin-film compounds. In the two-magnon scattering process, the uniform FMR mode (\textbf{k}=0) can scatter to degenerate non-uniform modes (\textbf{k}$\neq$ 0) with an effective interaction matrix proportional to the Fourier transform of samples inhomogeneities\cite{Heinrich}. We observe that the role of the scattering centres changes at low temperatures in \mgctwo where a superposition of twofold and fourfold symmetry dominates and not the usual six-fold as at room temperature. The ferromagnetic resonance measurements demonstrate the very good structural quality and the low magnetic damping in the C-doped \mg, paving the way for heterostructure integration and spintronic applications.

\section{Experimental details}


The sample preparation as well as the reflection high-energy electron diffraction (RHEED) measurements were performed in a UHV setup with a base pressure of 2.7$\times 10^{-8}$ Pa. \mgx layers were grown epitaxially on Ge(111) substrates\cite{Zeng2003,Mendez2008}.  These substrates were chemically cleaned before introduction in the UHV chamber. Then we did a degassing of the Ge(111) substrates by direct heating up to 720 K for 12 h and flashed afterwards at 1020 K to remove the native oxide layer. After repeated flashes at 1020 K and a cooling down at 770 K, a 15 nm thick Ge buffer layer was deposited on the Ge(111) substrates to make sure that the starting surface of the \mgx thin films growth is of good quality. The quality of this starting surface was checked \emph{in-situ} by RHEED. Eventually the sample was cooled down to room temperature (RT).

To form the \mgx layers we used the reactive deposition epitaxy method\cite{Petit2015}.  Using this method the \mgx layers are created by phase nucleation at the surface of the sample during the epitaxial growth. No diffusion phenomenon is required for the growth unlike the solid phase epitaxy process which is usually employed to form the \mgx films on Ge(111). However a good control of the different flows is needed to match the stoichiometry of the desired compound : Ge and Mn were evaporated using Knudsen cells and C atomic flow was obtained thanks to a high purity pyrolytic graphite filament source (SUKO) from MBE-Komponenten. The Ge and Mn flows were calibrated with a water-cooled quartz crystal microbalance and the C flow was calibrated using the structure transition between the Si(001) (2$\times$1) and c(4$\times$4) reconstructions which occurs for a C deposited thickness of 0.4 atomic monolayer on Si(001) surfaces\cite{Simon2001}.  The growth of the \mgx films was monitored \emph{in-situ} by RHEED : the \mgx films growing epitaxially on a Ge(111) surface exhibit an easily identifiable RHEED ($\sqrt{3}\times\sqrt{3}$)R30$^\circ$ pattern which is characteristic of the \mg and \mgx compounds\cite{Mendez2008,Zeng2004a}.

\begin{figure}[bt!]
  \includegraphics[width=8cm]{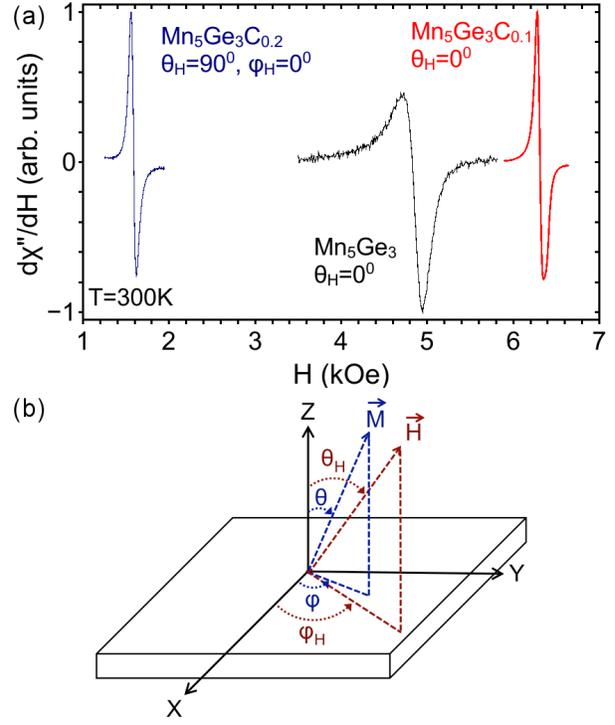}\\
 \caption{\label{Fig.0} (Color online) (a) Typical spectra at RT for 12nm thick \mg, \mgcone and \mgctwo thin films. (b) Schema of the coordinate system used in the FMR measurements.}
\end{figure}

The saturation magnetisation and the estimated Curie temperatures of all samples were determined by SQUID measurements. A SQUID magnetometer Quantum Design MPMSXL working in a temperature range 1.8K to 300K and in a magnetic field up to 5T was used. The FMR measurements were performed with a conventional X-band (9.39GHz) Bruker EMX spectrometer in the 80K to 300K temperature range. The samples ($2\times2 $mm$^2$) were glued on a quartz suprazil rode and mounted in the centre of a rectangular cavity (TE$_{102}$). To improve the signal-to-noise ratio, the FMR measurements were carried out using a modulation field of 100kHz and 5Oe amplitude with a lock-in detection. The FMR spectra were measured with the applied magnetic field rotated in plane and out-of-plane. The FMR spectra were fitted to a Lorentzian profile and the resonance field and FWHM linewidth were subsequently extracted. Typical spectra at RT are shown in figure~\ref{Fig.0}(a) for 12nm thick films. The signal-to-noise ratio (SNR) of \mg at 300K (figure~\ref{Fig.0}) is 38 which is the lowest SNR of all measured spectra and is due to the proximity to T$_c$. The SNRs for the other spectra of figure~\ref{Fig.0} are 115 and 250 for \mgctwo and \mgcone respectively which have higher T$_c$.

\section{Model and geometry}

The FMR spectra were analyzed with the Smit-Beljers formalism for a thin film with uniaxial (hexagonal) symmetry\cite{Farle}. For a ferromagnetic film with hexagonal symmetry, the free energy density including the Zeeman energy, the demagnetizing energy and the anisotropy energy density is written as:

\begin{eqnarray}\label{eq1}
F &= -MH[ \sin\theta\sin\theta_H\cos(\varphi-\varphi_H) + \cos\theta\cos\theta_H ] \nonumber\\
&- (2\pi M^2 - K_2)\sin^2\theta + K_4\sin^4\theta + K_{6\perp}\sin^6\theta \nonumber\\
&+ K_{6\parallel}\sin^6\theta\cos6\varphi
\end{eqnarray}

\noindent where $\theta_H$, $\varphi_H$ are the polar and azimuthal angle of the external field with respect to the surface normal of the thin film (the [001] direction) and, respectively, the [100] direction, $\theta$ and $\varphi$ are the polar and azimuthal angle of the magnetisation with respect same directions (figure~\ref{Fig.0}(b)) and K$_i$ are the anisotropy constants to sixth order. The resonance condition, neglecting the damping effects and considering the magnetisation at equilibrium under steady field, is given by:

\begin{equation}\label{eq2}
\Big( \frac{\omega}{\gamma} \Big) ^2 = H_1\cdot H_2 
\end{equation}

\noindent where $H_{1}$ and $H_2$ represent the stiffness fields evaluated at the equilibrium angles of the magnetisation:

\begin{eqnarray}
H_1  = \frac{1}{M} \frac{\partial^2 F}{\partial \theta ^2} \label{eq3}\\ 
H_2  = \frac{1}{M \sin ^2 \theta} \frac{\partial^2 F}{\partial \varphi ^2} \label{eq4}
\end{eqnarray}

Equation \eqref{eq2} is valid for a high-symmetry case, where the mixed second derivative of the free energy is nil. Our experiments were carried out in two distinct geometries:

\begin{enumerate}[label=(\roman*)]
\item out-of-plane geometry ($\varphi_H = 0^{\circ}$, $\theta_H$ variable). The stiffness fields are the following:
\begin{eqnarray}
& H_1^{out} = H_r\cos(\theta-\theta_H) - 4\pi M_{eff}\cos 2\theta  \nonumber\\
& +2\frac{K_4}{M}(\cos 2\theta -\cos 4\theta) + 30\frac{(K_{6\perp}+K_{6\parallel})}{M}\sin^4\theta \nonumber\\
& -36\frac{(K_{6\perp}+K_{6\parallel})}{M}\sin^6\theta \label{eq5} \\ 
& H_2^{out} = H_r\cos(\theta-\theta_H) - 4\pi M_{eff}\cos^2\theta \nonumber\\ 
& +4\frac{K_4}{M}(\cos^2\theta - \cos^4\theta) + 6\frac{(K_{6\perp}+K_{6\parallel})}{M}\sin^4\theta\cos^2\theta \nonumber\\ & - 36\frac{K_{6\parallel}}{M}\sin^6\theta  \label{eq6}
\end{eqnarray}

\item in-plane geometry ($\theta_H = 90^{\circ}$, $\varphi_H$ variable). The stiffness fields are:

\begin{eqnarray}
& H_1^{in} = H_r\cos(\varphi-\varphi_H) + 4\pi M_{eff} -4\frac{K_4}{M} - 6\frac{K_{6\perp}}{M} \nonumber\\
& - 6\frac{K_{6\parallel}}{M}\cos6\varphi  \label{eq7} \\
& H_2^{in} = H_r\cos(\varphi-\varphi_H) - 36\frac{K_{6\parallel}}{M}\cos6\varphi \label{eq8}
\end{eqnarray}
\end{enumerate}

Here $4\pi M_{eff} =  4\pi M - 2K_2/M$, $\omega$ the angular frequency and $\gamma$ = g$\mu_B$/$\hbar$ the gyromagnetic ratio.

The FMR linewidth is analyzed by including the intrinsic and extrinsic damping mechanisms\cite{Chappert,Platow,Mizukami} :

\begin{equation}\label{eq9}
\Delta H = \Delta H_{intr} + \Delta H_{extr}
\end{equation}

In this expression, the intrinsic contribution due to the magnon-electron interaction can be described by the dimensionless Gilbert damping parameter $\alpha$\cite{Gilbert,Zakeri}:

\begin{equation}\label{eq10}
\Delta H_{intr} = \frac{2\alpha\omega}{\gamma\Psi}
\end{equation}

\noindent where $\Psi =\frac{1}{H_1 + H_2}\frac{d(\omega^2/\gamma^2)}{dH_r}$ is the dragging function as the magnetisation \textbf{M} is dragged behind \textbf{H} owing to anisotropy. When \textbf{M} and \textbf{H} are parallel, this contribution vanishes. As generally the in-plane and out-of-plane linewidth are not equal, extrinsic contribution have to be taken into account. The extrinsic contribution generally include the magnetic relation due to magnon-magnon interaction, the two-magnon interaction, which is given by\cite{Arias,Landeros,McMichael,Kalarickal}:

\begin{equation}\label{eq11}
\Delta H_{2mag} = \frac{\Gamma}{\Psi}
\end{equation}

\noindent with $\Gamma$ the two-magnon scattering rate. The two-magnon contribution usually vanishes for a critical out-of-plane angle $\theta < 45^{\circ}$. Short-range fluctuations in magnetic properties due to sample imperfections lead to two-magnon scattering. Inhomogeneous broadening effects (superposition of local resonances) due to long-range fluctuations of magnetic properties also participate to the extrinsic linewidth, especially at intermediate angles as the resonance local field can vary. We consider here three types of inhomogeneous broadening: $\Delta H_{mos}, \Delta H_{int}$ and $\Delta H_{inhom}$. The first term is the mosaicity term due to the distribution of easy axes directions\cite{Chappert,Zakeri}:

\begin{equation}\label{eq12}
\Delta H_{mos} = \left\vert \frac{\partial H_r}{\partial \beta_H} \right\vert \Delta \beta_H
\end{equation}

\noindent with $\beta_H$ = ($\theta_H,\varphi_H$). The second term represents the inhomogeneity of the internal fields in the sample\cite{Mizukami}:

\begin{equation}\label{eq13}
\Delta H_{int} = \left\vert \frac{\partial H_r}{\partial (4\pi M_{eff})} \right\vert \Delta (4\pi M_{eff})
\end{equation}

Finally, the last term which can contribute to the linewidth is a residual frequency and angular independent inhomogeneous linewidth that cannot be put in other form and is also due to long-range fluctuations of magnetic properties.

The procedure used to determine the magnetic parameters was as follows: the anisotropy fields were determined using the system of equations \eqref{eq5}-\eqref{eq8} applied at high symmetry directions (along easy/hard axes) together with the corresponding measured resonance fields (fixed frequency) at a fixed \textit{g}-factor. Afterwards, the polar and azimuthal angular dependence of the resonance field was fitted to the same equations and the equilibrium condition of the free energy allowing for a variable \textit{g}-factor as parameter. The iteration was repeated several times until a good fit was obtained. This analysis yields the \textit{g}-factor, the anisotropy constants and the magnetisation direction $\theta$. These values serve in the fit of the angular variation of the linewidth which allows the numerical evaluation of $\alpha$, $\Gamma$ and the inhomogeneous contributions using  \eqref{eq2}-\eqref{eq13}.


\section{Results and discussion}

In this section, experimental results of C-doped \mg thin films investigated by ferromagnetic resonance and SQUID magnetometry are presented. Using samples with different carbon content, we determined the magnetic anisotropy energy, the \textit{g}-factor, magnetisation and magnetic relaxation parameters.

\subsection{Magnetic anisotropy}

\begin{figure*}[ht!]
  \includegraphics[width=13cm]{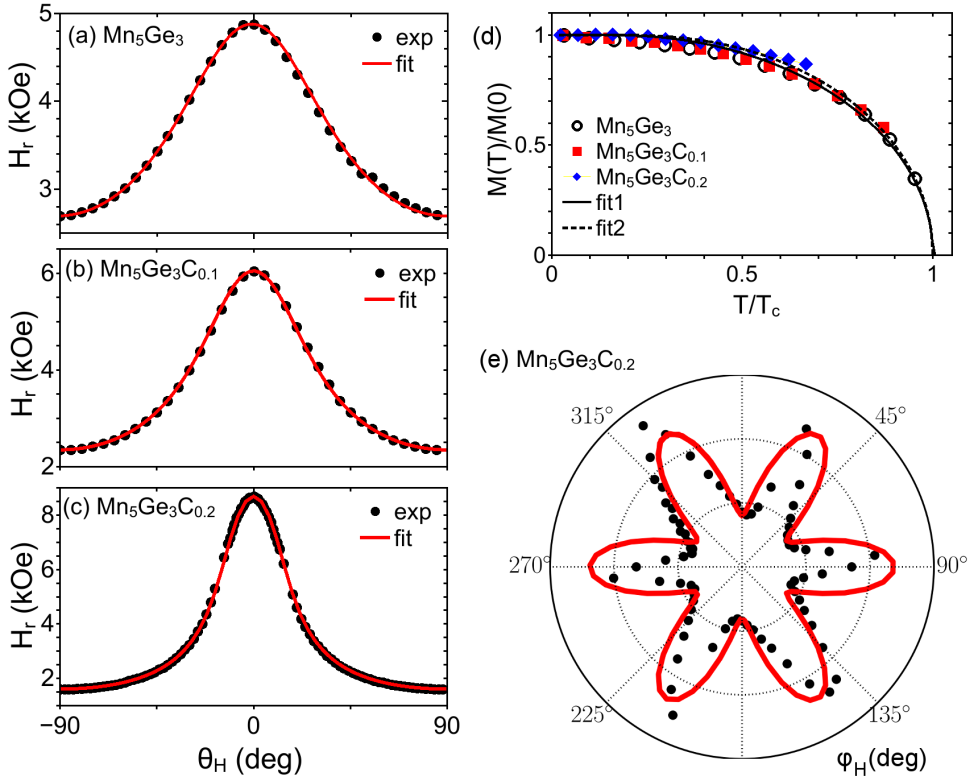}\\
 \caption{\label{Fig.1} (Color online) Out-of-plane angular variation of the resonance field at 300K for (a) \mg, (b) \mgcone, (c) \mgctwo. The temperature dependence of the magnetisation is shown in (d) in normalised coordinates. The full and dotted lines correspond to fits with a Brillouin function. The estimated T$_c$s are 315K, 345K and 450K. (e) In-plane angular dependence of the resonance field for \mgctwo at room temperature. The diagram centre and the dotted circles correspond to H$_r$ values of 1605, 1606 and 1607 Oe respectively. The error bars are $\pm$ 0.07 Oe. The line represents a fit to \eqref{eq2}.}
\end{figure*}

To determine the magnetic energy anisotropy (in absolute units), FMR measurements were carried out at a frequency of 9.4GHz. The FMR spectra were recorded as a function of the polar and azimuthal angles of the external magnetic field at different temperatures. The saturation magnetisation was determined from SQUID measurements. In figure~\ref{Fig.1}(d), the temperature dependence of the magnetisation up to 300K is shown for \mg, \mgcone and \mgctwo. The Curie temperature was estimated from these curves using a mean-field approximation which gives a good approximation of the temperature dependence of the magnetisation. The full line correspond to a fit with the Brillouin function B$_{1.5}$ (S=3/2) and the dotted line to a fit with B$_1$ (S=1) that follows better the experimental points of \mgctwo. The estimated values of T$_c$ are 315K, 345K and 450K. The error bars correspond to $\pm$10K for \mg and \mgcone as the experimental points cover a larger temperature range and superpose closely with B$_{1.5}$. The experimental points for \mgctwo cover only a small part of the temperature range and the error bars are estimated to be of $\pm$30K.

The out-of-plane angular variation for the resonance field H$_r$ is shown in figure~\ref{Fig.1}(a)-(c) for \mg, \mgcone and \mgctwo at room temperature. The H$_r$($\theta_H$) curves indicate an in-plane easy axis with a minimum resonance field of 1.6kOe, 2.3kOe and 2.7kOe for \mgctwo, \mgcone and \mg respectively for $\varphi_H=0^{\circ}$. The hard axis is perpendicular to plane ([001] direction) and has the highest H$_r$ of 8.6kOe, 6kOe and 5kOe. The azimuthal angular dependence of the resonance field for \mgctwo, recorded also at 300K is shown in figure~\ref{Fig.1}(e). The sixfold (hexagonal) symmetry in the azimuthal angular dependence indicates that an in-plane hexagonal anisotropy exists with one easy axis along the [100] direction of the film. The experimental FMR data of out-of-plane and in-plane dependence of the resonance field can be well simulated with \eqref{eq2} and the anisotropy fields can be extracted. The anisotropy constants can be found in absolute units by using the sample magnetisation determined from SQUID measurements.

\begin{table*}[!h]
\caption{\label{Table1}Magnetic parameters for \mg, \mgcone and \mgctwo at different temperatures obtained from the FMR.}
\begin{tabular*}{\textwidth}{@{\extracolsep{\fill} } l| *{6}{c}}
\br
Sample & T(K) & 4$\pi$ M$_{eff}$(kOe) & K$_2$(erg/cm$^3$) & K$_4$(erg/cm$^3$) & K$_{6 \parallel}$(erg/cm$^3$) & $\gamma/2\pi$(GHz/kOe) \\ \mr
\multirow{3}{*}{\mg} & 300 & 1.5 & 0.37$\times$ 10$^6$ & 2832 &  & 2.8\\
& 200 & 4.0 & 2.01$\times 10^6$ & -2.28$\times 10^5$ & & 2.8\\
& 150 & 4.8 & 2.36$\times 10^6$ & -2.93$\times10^5$ & & 2.8\\ \mr
\multirow{4}{*}{\mgcone} & 300 & 2.6 & 1.65$\times$ 10$^6$ & 3.85$\times 10^4$ &  & 2.8\\
& 250 & 3.8 & 2.71$\times$ 10$^6$ & -1901 & & 2.8\\
& 200 & 4.4 & 3.37$\times 10^6$ & -5131 & & 2.8\\
& 100 & 5.0 & 4.29$\times 10^6$ & 2.58$\times 10^4$ & & 2.8\\ \mr
\multirow{4}{*}{\mgctwo} & 300 & 5.3 & 4.39$\times$ 10$^6$ & 4.41$\times 10^4$ & 27 & 2.84\\
& 250 & 5.8 & 4.78$\times$ 10$^6$ & 5.53$\times 10^4$ & 134 & 2.84\\
& 150 & 6.6 & 5.19$\times$ 10$^6$ & 5.35$\times 10^4$ & & 2.84\\
& 100 & 7.0 & 5.28$\times$ 10$^6$ & 4.61$\times 10^4$ & & 2.84\\ \br
\end{tabular*}
\end{table*}

The resulting anisotropy constants are summarised in Table~\ref{Table1} along with the \textit{g}-factor at several temperatures. The positive sign of K$_2$ indicates that this term favors an out-of-plane easy axis of magnetisation while the shape anisotropy dominates\cite{Truong}. In the very thin film limit, K$_2$ could overcome the shape anisotropy resulting in an out-of-plane anisotropy axis. The different K$_i$ have a different temperature dependence. For \mg and \mgcone, the sixfold in-plane symmetry is to low to be extracted, therefore only the K$_2$ and K$_4$ constants were determined from the angular measurements. K$_2$ is positive for \mg and C-doped \mg at all temperatures and increases at low temperature. K$_4$ decreases (increases in absolute values) for \mg, but for the C-doped compounds has a minimum or a maximum at an intermediate temperature. The sixfold in-plane anisotropy in \mgctwo increases at 250K from the room temperature value, while at lower temperature becomes to small or a transition to a fourfold in-plane anisotropy arises as will be inferred from the linewidth temperature dependence discussed in the next section. The error bars for the anisotropy constants presented in Table~\ref{Table1} are around 5$\%$, which come from the computation of M from SQUID measurements due to the uncertainty in the volume of our samples.

The \textit{g}-factor can be estimated from the angular dependence of the resonance field. Its value indicates the influence of the orbital contribution to the total magnetic moment. The ratio of the orbital to the spin magnetic moment can be inferred from the Kittel formula and is equal to the deviation of the \textit{g}-factor from the free electron value. The value of the \textit{g}-factor for \mg and \mgcone is 2.0005, while for \mgctwo this value increases to 2.0291 meaning an increased orbital contribution with carbon doping (1.5\% of the spin magnetic moment).

\subsection{Magnetic relaxation}

\begin{figure}[t!]
  \includegraphics[width=8.5cm]{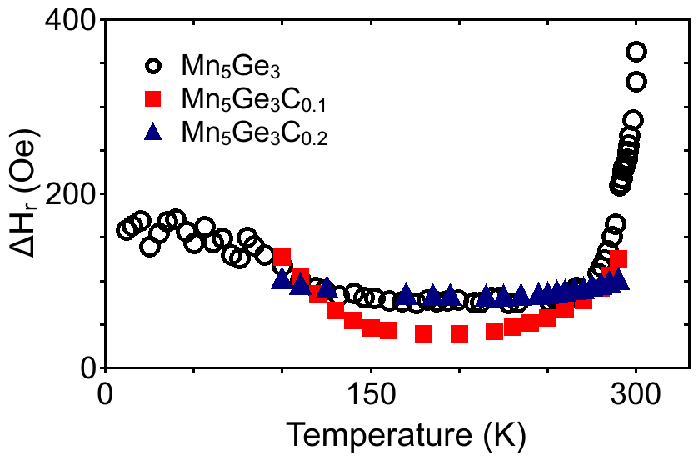}\\
 \caption{\label{Fig.2} (Color online) Temperature variation of the resonance linewidth for \mg, \mgcone and \mgctwo.}
\end{figure}

The linewidth of the resonant signal $\Delta H_r$ is directly related to the magnetic and structural quality of the films and provide information about the different relaxation channels in magnetic damping. In figure~\ref{Fig.2}, the temperature dependence of the FMR linewidth is shown for the perpendicular to plane direction ($\theta_H=0^{\circ}$) for \mg and C-doped \mg. A shallow minimum is observed for all three compounds around 200K and a sharp peak close to T$_c$. At lower temperature, the FMR linewidth increases and saturates for \mg (measured to 6K). The minimum in the linewidth seems not related with the C-doping. It occurs around the same absolute value of temperature and could be related with a small in-plane transition to a fourfold anisotropy from sixfold anisotropy (tetragonal distortion) or to a constriction by the substrate. The increase of linewidth at low temperature was explained as an inhomogeneous broadening due to the increase of the anisotropy constants (K$_2$) with decreasing temperature\cite{Platow}. The error bars for the measured FMR linewidth are less than 2$\%$ in all cases.

\begin{figure}[!th]
  \includegraphics[width=7cm]{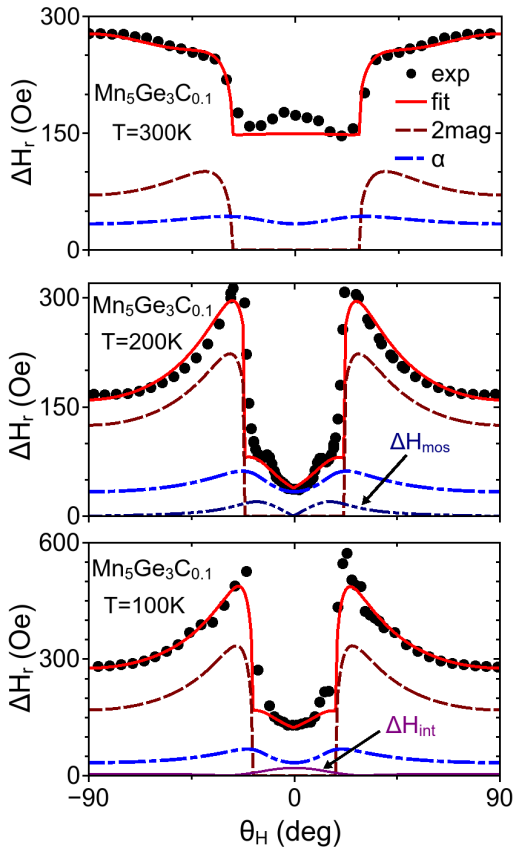}\\
 \caption{\label{Fig.3} (Color online) Out-of plane angular dependence of the resonance linewidth for \mgcone at different temperatures. The full lines represent a fit with intrinsic and extrinsic contributions, the dashed lines indicate the two-magnon contribution while the dash-dotted lines indicate the intrinsic contribution. The symbols indicate experimental data. The typical mosaicity and internal fields ($\Delta(4\pi M_{eff})$) contributions are shown at 200K and 100K respectively.}
\end{figure}

\begin{figure*}[!h]
  \includegraphics[width=17.5cm]{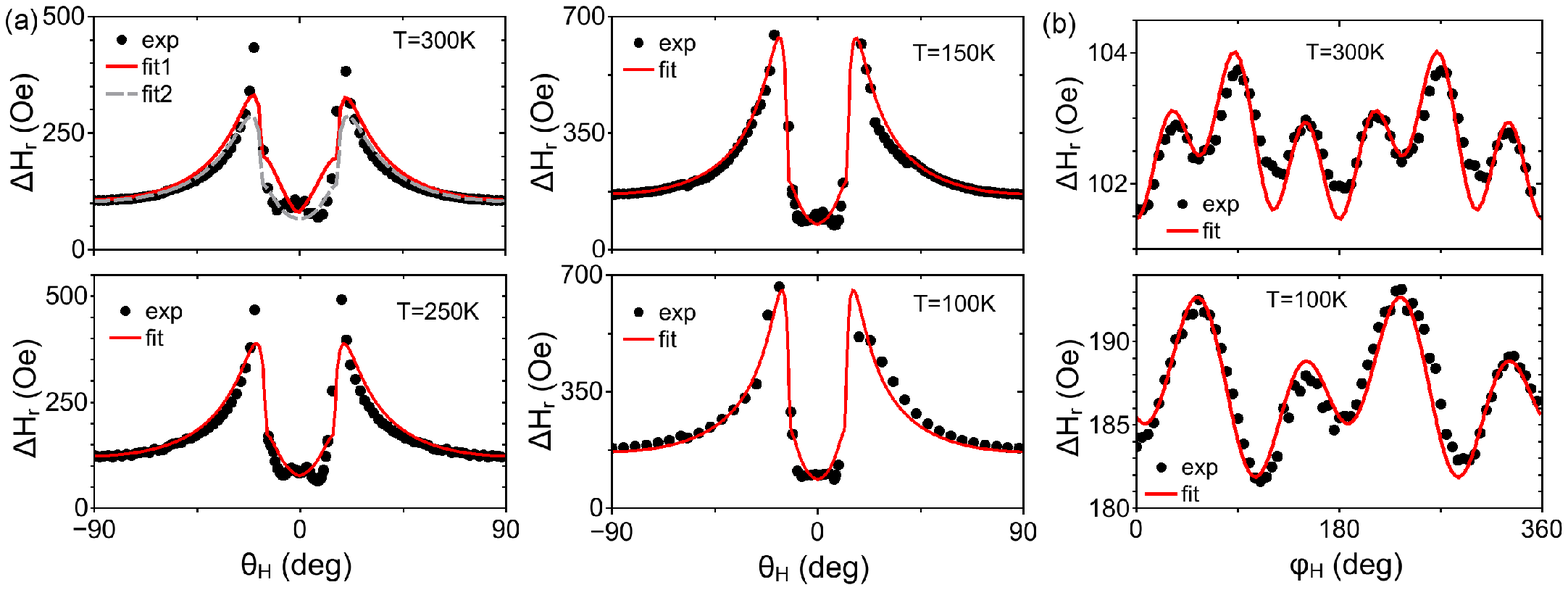}\\
 \caption{\label{Fig.4} (Color online) Out-of plane (a) and in-plane (b) angular dependence of the resonance linewidth for \mgctwo at different temperatures. The lines represent fits with intrinsic and extrinsic contributions. The error bars in panel (b) are $\pm$0.2 Oe at 300K and $\pm$0.8 Oe at 100K.  }
\end{figure*}

Figure~\ref{Fig.3} and figure~\ref{Fig.4}(a) show the out-of-plane variation of the FMR linewidth for the C-doped \mg at room and low temperatures. The shape of the curves shows the characteristic dependence for thin films with a maximum of the linewidth at intermediate angles. Our films have an in-plane easy axis at all temperatures, therefore the magnetisation lags behind the applied field when the field is rotated out of the plane. The peak in the linewidth occurs for  $\theta_H$ between 20$^{\circ}$ at room temperature and 10$^{\circ}$ at low temperature, corresponding to the largest interval between \textbf{M} and \textbf{H}. From the theoretical fits of the data (solid lines) obtained using \eqref{eq2}-\eqref{eq13}, the relaxation parameters are extracted and listed in Table~\ref{Table2}.

For all compounds, the perpendicular to plane linewidth is always smaller than the in-plane one indicating the presence of two-magnon scattering and other extrinsic contributions in the samples. The intrinsic damping alone cannot explain the out-of-plane shape of the $\Delta H_r(\theta_H)$ dependence as observed in figure~\ref{Fig.3}, where the two-magnon and intrinsic Gilbert contribution are shown individually for comparison. The respective values of the two contributions are further compared in Table~\ref{Table2} by computing the Gilbert damping parameter G and the two-magnon scattering contribution $\gamma\Gamma_{2mag}$ in s$^{-1}$. The contribution of the intrinsic linewidth (Gilbert) to the total linewidth varies between 7$\%$ and 33$\%$. The mosaicity contribution is low for all cases and its angular dependence induces the V-shaped form around $\theta_H$=0$^{\circ}$. A typical mosaicity contribution is shown in figure~\ref{Fig.3} for \mgcone at 200K. The inhomogeneity of the internal field gives mainly a low contribution around $\theta_H$=0$^{\circ}$, with the exception of \mgcone at room temperature where it gives a larger contribution for $\theta_H$ close to the plane. A typical variation of this contribution is shown in figure~\ref{Fig.3} at 100K.

The estimated intrinsic damping $\alpha$ is considered isotropic and independent of temperature in the considered temperature range (100K-300K). We prefer using the dimensionless parameter $\alpha$ which varies between 0.005 and 0.01 over the Gilbert damping parameter G given by $\alpha$=G/$\gamma$M as the latter will imply a strong temperature dependence. The Gilbert damping represents the decay of magnetisation by direct viscous dissipation to the lattice as it is introduced in the Landau-Lifschitz-Gilbert equation\cite{Gilbert}. The spin-orbit coupling is assumed to be at the origin of spin-lattice relaxation in ferromagnets. \textit{Ab-initio} calculations that include the spin-orbit coupling explicitly show a weak dependence of $\alpha$ with temperature in a large range of temperatures\cite{Gilmore,Ebert}. Two different mechanisms contribute to the temperature dependence\cite{Heinrich}, one conductivity-like and one resistivity-like with a transition between the two at intermediate temperature. Sometimes these two contributions have an equal influence on the damping. Other models predict an anisotropic tensorial damping constant $\hat{\alpha}$\cite{Safonov} and its proportionality to 1/M\cite{Vittoria}, leading to a strong variation close to T$_c$.

We estimated the value of $\alpha$ for each compound by fitting the out-of-plane angular dependence of $\Delta H_r$ at a temperature corresponding to the minimum of the curves in figure~\ref{Fig.2} (around 200K). For this specific temperature, the estimation corresponds to the maximum possible value of $\alpha$ considering small inhomogeneous broadening ($\Delta H_{int}$ and $\Delta H_{inh}$). Although we consider a constant $\alpha$, as it is observed from Table~\ref{Table2}, at room and low temperature the linewidth (and correspondingly the inhomogeneous residual field) increases for \mgcone which could be explained by an increase of $\alpha$ at least at low temperature. The room temperature increasing in the linewidth is usually explained as a breakdown of the uniform precession due to thermal excitations (spin-waves)\cite{Li}. The increasing of the linewidth at low temperature is smaller for \mg and \mgctwo in the 100-300K temperature range being compatible with a constant $\alpha$ as considered for this temperature range.

\begin{table*}[!h]
\caption{\label{Table2}Magnetic relaxation parameters for \mg, \mgcone and \mgctwo determined from the out-of-plane angular variation of the FMR at different temperatures.}
\begin{tabular*}{\textwidth}{@{\extracolsep{\fill} } l| *{8}{c}}
\br
Sample & T(K) & $\alpha$ & G(10$^8$s$^{-1}$) & $\Gamma_{2mag}$(Oe) & $\gamma \Gamma_{2mag}$(10$^8$s$^{-1}$) & $\Delta\theta_H$(deg) & $\Delta (4\pi$M$_{eff}$)(Oe) & $\Delta H_{inh}$(Oe) \\ \mr
\multirow{3}{*}{\mg} & 300 & 0.01 & 0.54 & 150 & 26.38 & 0.05 & 20 & 270\\
& 200 & 0.01 & 1.29 & 340 & 59.81 & 0.1 & 10 & 70 \\
& 150 & 0.01 & 1.41 & 550 & 96.76 & 0.1 & 10 & 10 \\ \mr
\multirow{5}{*}{\mgcone} & 300 & 0.005 & 0.55 & 210 & 36.94 & 0.05 & 80 & 80\\
& 250 & 0.005 & 0.73 & 280 & 49.26 & 0.1 & 5 & 15 \\
& 200 & 0.005 & 0.81 & 320 & 56.29 & 0.1 & 5 & 5\\
& 150 & 0.005 & 0.88 & 400 & 70.37 & 0.1 & 10 & 5\\
& 100 & 0.005 & 0.92 & 430 & 75.64 & 0.1 & 15 & 70\\ \mr
\multirow{4}{*}{\mgctwo} & 300 & 0.01 & 1.92 & 220 & 39.25 & 0.05-0.2 & 10 & 5\\
& 250 & 0.01 & 2.02 & 300 & 53.53 & 0.05 & 10 & 5\\
& 150 & 0.01 & 2.16 & 500 & 89.22 & 0.05 & 10 & 5\\
& 100 & 0.01 & 2.21 & 450 & 80.30 & 0.05 & 10 & 5\\ \br
\end{tabular*}
 \begin{tablenotes}
      \small
      \item \textit{Note}: The error bars are around 10$\%$.
    \end{tablenotes}
\end{table*}

The second relaxation mode that influence the FMR linewidth is the two magnon scattering. The uniform mode can couple with degenerate spin-wave modes due to fluctuations in the local effective field that can arise from surface defects, scattering centres, fluctuation in the anisotropy from grain to grain or other inhomogeneities\cite{Arias,McMichael}. The two magnon scattering rate $\Gamma$ depends on the angle $\theta_H$ (out-of-plane geometry) and on the resonance field H$_{res}$. A detailed analysis based on the effect of the defects on the response functions of thin films was performed in \cite{Landeros} and \cite{Lindner} for the case when the magnetisation is tipped out-of-plane. We consider here the same type of angular dependence of $\Gamma$ as in \cite{Lindner} (see (8)). $\Gamma$ depends on the nature and shape of the defects that activate the scattering mechanism. The values for the \mg compounds, extracted from the fitting of the linewidth curves, are shown in Table~\ref{Table2} as a function of temperature. From the calculated value $\Gamma_{2mag}$=8H$_K b^2 p/\pi D$, the exchange spin-wave stiffness D can be inferred if details of the defects  are known such as the covered fraction of the surface $p$ or the effective height $b$ (H$_K$ the anisotropy field). Atomic force microscopy measurements were performed on the samples, from which the rms surface roughness was determined: for \mg the surface roughness was about 1.5-2nm, while for \mgx it was close to 1nm. Therefore, at room temperature, the spin-wave stiffness was estimated as 0.12$\times 10^{-8}$G cm$^2$ for \mg, 0.16$\times 10^{-8}$G cm$^2$ for \mgcone and 0.39$\times 10^{-8}$G cm$^2$ for \mgctwo considering a defect ratio of 50$\%$. These values are only estimates as a precise identification of the defects is difficult to obtain.

As observed from Table~\ref{Table2}, the other extrinsic contributions to the linewidth have only a small impact on the fitted curves. The mosaicity is very small, inferior to 0.1$^{\circ}$, being a testimony of the good quality of our samples. Also the inhomogeneity of the internal fields is almost negligible in the majority of cases, only for \mgcone at room temperature it seems to have a larger influence. The higher values of H$_{int}$ are needed to explain the small peak observed around $\theta_H$ = 0$^{\circ}$ for both \mg and \mgcone and for the increase of the linewidth at intermediate angles until $\theta_H$ = 90$^{\circ}$ for \mgcone at room temperature. The values of the residual inhomogeneous contribution are generally small, the larger values can also be attributed to a temperature dependent intrinsic contribution as discussed above.

We now discuss the case of \mgctwo, for which both out-of-plane and in-plane data were fitted, as shown in figure~\ref{Fig.4}. The panel (a) show the out-of-plane dependence of the FMR linewidth. The 300K and 250K data are well fitted close to $\theta_H$ = 0$^{\circ}$ and at larger angles but not at the peaks that correspond to the largest interval between \textbf{M} and \textbf{H} (critical angle). The dashed line at T=300K corresponds to a fit with the parameters indicated in Table~\ref{Table2} and $\Delta\theta_H$ = 0.05$^{\circ}$, while the full line to a fit with $\Delta\theta_H$ = 0.2$^{\circ}$. Although increasing the mosaicity contribution fits better the peaks, the  fitted curve becomes V-shaped between the peaks in total contradiction with the data. We believe that the mosaicity is small (0.05$^{\circ}$) and the discrepancy at the critical angle at 300K is due to some other effect (the FMR line being strongly distorted at this angle). We also tried to fit the 300K curve introducing in-plane second and fourth order anisotropy constants (K$_{2\parallel}$ and K$_{4\parallel}$) without a better result (not shown). The low temperature curves are nicely fitted with the presented model for all angles. 

\begin{table*}
\caption{\label{Table3}Magnetic relaxation parameters for \mgctwo at different temperatures determined from the in-plane angular variation of FMR.}
\begin{tabular}{ *{8}{c}}
\br
T(K) & $\Gamma_{0}$(Oe) & $\Gamma_{2}$(Oe) & $\Gamma_{4}$(Oe) & $\Gamma_{6}$(Oe) & $\varphi_2$ & $\varphi_4$ & $\varphi_6$ \\ \mr
300 & 72.75 & 1.5 &  & 1.5 & 90 &  & 30\\
250 & 97.5 & 1.7 &  & 1.5 & 90 &  & 30\\
150 & 254.2 & 8.6  & 5.5 &  & 57 & 166 & \\
100 & 291.4 & 12.4 & 8.6 &  & 57 & 167 & \\ \br
\end{tabular}
\end{table*}

For the in-plane dependence of $\Delta H_r$, the only contributions that were considered were from the isotropic intrinsic damping and the two-magnon contribution which was expressed as follows\cite{Arias,Zakeri,Lindner}:

\begin{eqnarray}
\Delta H_{2mag} = \frac{\sum_i \Gamma_i f(\varphi_i)}{\Psi} \times \nonumber\\ \arcsin \left( \sqrt{\frac{\sqrt{\omega_r^2+(\omega_0 /2)^2}-\omega_0/2}{\sqrt{\omega_r^2+(\omega_0/2)^2}+\omega_0/2}}\right)
\end{eqnarray}

\noindent with $\omega_0=\gamma M_{eff}$ and $\Gamma_i f(\varphi_i)$ characterise the anisotropy of the two-magnon scattering along different crystallographic in-plane directions. At 300K and 250K (figure~\ref{Fig.4}(b)), the FMR linewith has the same six-fold symmetry as the angular dependence of $H_r$ (figure~\ref{Fig.1}(e)). If the scattering centres are given by lattice defects (dislocation lines), the azimuthal dependence should reflect the lattice symmetry\cite{Woltersdorf,Zakeri}. The angular dependence of the scattering was fitted with $\Gamma_i f(\varphi_i) = \Gamma_0 + \Gamma_2 \cos^2(\varphi-\varphi_2) + \Gamma_6\cos 6(\varphi-\varphi_6)$ at 250K and 300K and with $\Gamma_i f(\varphi_i) = \Gamma_0 + \Gamma_2 \cos^2(\varphi-\varphi_2) + \Gamma_4\cos 4(\varphi-\varphi_4)$ at 150K and 100K. The parameters $\Gamma_2$ and $\Gamma_4$ are phenomenologically introduced to account for the observed angular variation. $\Gamma_6$ is expected from the sixfold symmetry. The in-plane anisotropies are very small as observed form their values in Table~\ref{Table1}, therefore $\varphi_M\approx\varphi_H$ and the dragging function is very close to unity and neglected. A change of symmetry of the scattering (or of the dominant contribution to the scattering) seems to take place around 200K corresponding to the minimum in figure~\ref{Fig.2}. At lower temperature a superposition of twofold and fourfold symmetry dominates the angular dependence of the in-plane linewidth. This can be related to the shape and orientation of the defects (rectangular) that cause the two-magnon scattering\cite{Arias,Woltersdorf,Lenz}. The two-magnon scattering intensity depends on the direction of the magnetisation in respect to the symmetry axes of the magnetic defects and on the angle between the magnetisation and the crystallographic axes. One can think that the twofold and fourthfold symmetry can be related to the magnetic symmetry of the defects and to the lattice defect symmetry at low temperature with the defects oriented mostly along the [110] direction. This is also consistent with the increase in the two-magnon scattering rate (see Table~\ref{Table2} and Table~\ref{Table3}) at lower temperature and with the variation in the two-magnon linewidth contribution to the total linewidth (from around 30$\%$ at 300K to 66$\%$ at 100K).


\section{Conclusion}

12nm thick \mg and \mgx films were grown by reactive deposition epitaxy on Ge(111) substrates. Detailed FMR measurements were performed on the samples at different temperatures. Both \mg and C-doped \mg show perpendicular uniaxial magneto-crystalline anisotropy and an in-plane easy axis of magnetisation due to the large shape anisotropy. Doping with carbon increases the anisotropy fields without diminishing the quality of the sample as demonstrated by the small FMR linewidth of the films. From the angular dependence of the resonance field and of the linewidth, the anisotropy fields, \textit{g}-factor and magnetic relaxation parameters are obtained. The contributions to the broadening of the FMR linewidth come primarily from the intrinsic Gilbert damping and two-magnon scattering. In the temperature dependence of the out-of-plane linewidth a very shallow and wide minimum is observed close to 200 $\pm$ 40K. One can suppose that this minimum and the changing of the dominating contribution in the two-magnon scattering from centres with different symmetry are connected one to each other. Nevertheless this supposition requires additional evidences.


\ack

This work has been carried out thanks to the support of the A*MIDEX project (No. ANR-11-IDEX-0001-02) funded by the "Investissements d'Avenir" French Government program, managed by the French National Research Agency (ANR). This work is part of A*MIDEX through the HIT project APODISE. We also want to thank the interdisciplinary French EPR network RENARD (CNRS - FR3443). 


\end{document}